\documentclass{ifacconf}

\usepackage{graphicx}      
\usepackage{amssymb,amsmath,amsfonts,mathtools}
\usepackage{natbib}        
\usepackage{xcolor}
\usepackage{algorithm}
\newtheorem{assumption}{Assumption}
\newtheorem{problem}{Problem}

\begin{document}
\begin{frontmatter}

\title{Model fusion for efficient learning of nonlinear dynamical systems} 


\author[First]{Vatsal Kedia} 
\author[Second]{Vivek S. Pinnamaraju} 
\author[Third]{Dinesh Patil} 

\address[First]{Electrical Engineering Department, Indian Institute of Technology Bombay, Mumbai, India (e-mail: vatsalkedia@ ee.iitb.ac.in).}
\address[Second]{ABB Corporate Research, Bangalore, India \\ (e-mail: vivek.pinnamaraju@ in.abb.com)}
\address[Third]{ABB Corporate Research, Bangalore, India \\ (e-mail: dinesh.patil@ in.abb.com)}

\begin{abstract}                
In the context of model-based control of industrial processes, it is a common practice to develop a data-driven linear dynamical model around a specified operating point. However, in applications involving wider operating conditions, representation of the dynamics using a single linear dynamic model is often inadequate, requiring either a nonlinear model or multiple linear models to accommodate the nonlinear behaviour. While the development of the former suffers from the requirements of extensive experiments spanning multiple levels, significant compromise in the nominal product quality and dealing with unmeasured disturbances over wider operating conditions, the latter faces the challenge of model switch scheduling and inadequate description of dynamics for the operating regions in-between. To overcome these challenges, we propose an efficient approach to obtain a parsimonious nonlinear dynamic model by developing multiple linear models from data at multiple operating points, lifting the data features obtained from individual model simulations to adequately accommodate the underlying nonlinear behaviour and finally, sparse optimization techniques to obtain a parsimonious model. The performance and effectiveness of the proposed algorithm is demonstrated through simulation case studies. 
\end{abstract}


\end{frontmatter}

\section{Introduction}

Data-driven learning of linear dynamical systems is very popular owing to its ease of interpretation, strong theoretical guarantees, mathematical tractability and wide range of applications (\cite{ljung1998system}; \cite{ljung2010perspectives}). In the context of process industries, linear dynamic models are easy to develop as they need simple experiments involving use of step tests or pseudo random binary signal (PRBS) excitation to generate informative data (\cite{zhu2001multivariable}). It is well known that linear models provide accurate approximation around an operating point but as we deviate away from the operating point the prediction becomes worse since the linearization approximation no longer holds (\cite{ren2022mltutorial}). Therefore linear models are not well suited for non-linear systems due to less flexibility and hence are unable to capture the all nonlinearities of the system. 

While it is reasonable to develop linear models around a particular operating point, a single linear model to capture the underlying dynamics is inadequate in applications involving wider operating regions. In order to ensure good control performance, it is necessary to describe the dynamics either using a nonlinear model or multiple linear models. For directly identifying nonlinear model we need input excitation covering the whole operating region of interest to have informative data since step and/or PRBS tests are inadequate (\cite{sriniwas1995nonlinear}). Hence we need to perform staircase or generalized multiple-level noise (GMN) test (\cite{zhu2001multivariable}) which disturbs the normal plant operation affecting the product quality and expensive due to prolonged test duration. There are several popular nonlinear model structures which can accommodate the nonlinear behavior such as NARX, NARMAX, Neural networks, etc. However, learning the underlying non-linear dynamics using these models poses significant practical challenges such as lack of parsimony, requirement of extensive experiments, complexity in parameter estimation etc. (see \cite{schoukens2019nonlinear} and references therein). Further, the neural networks approach to model dynamical systems includes recurrent neural networks (RNNs), LSTMs, etc. (see \cite{pillonetto2023deepnetwrok} and references therein for survey on the topic) necessitate larger datasets compared to first-principles (parsimonious) models. Moreover, these models often lack extrapolation capabilities and can be challenging to interpret due to non-parsimonious model structure. An additional limitation of these models lies in their mathematical structure, which poses challenges when deriving methods for analysis of closed-loop stability and performance (\cite{nikolakopoulou2020feedback}).

In this regard, multiple linear model framework to explain the nonlinear behavior over wider operating region has gained attention (\cite{johansen1999multiple}). For more details on this approach, the readers are encouraged to refer the review paper by \cite{adeniran2016modeling} and references therein. As the name suggests this approach involves developing different local linear models at different operating points. The models used for control are switched depending on the process operating condition. This approach faces several challenges such as deciding on number of local linear models to be developed, switching strategy for an intermediate operating region where none of the individual models are adequate to explain the underlying dynamics etc.  

Learning non-linear models poses several challenges as mentioned above. To overcome these challenges, we propose an efficient model fusion algorithm.
The proposed approach utilizes simulated data from locally learned linear models at different operating points, followed by lifting techniques to generate the features necessary to explain the nonlinear behavior and finally, sparse optimization techniques to obtain a parsimonious model. 
The key idea is while each model simulation generates locally linear data, the fusion across operating points essentially results in nonlinear data, which is appropriate for nonlinear model identification. Since the non-linearity of the underlying dynamic model is unknown, we postulate several non-linear features generated from the simulated data using the ideas of lifting. While this results in over-parameterized models, to obtain parsimonious models, we further use sparse regularization on the model parameters to enable feature selection and extract only the features that are most relevant in explaining the underlying nonlinear behavior (\cite{zou2005regularization}). Furthermore, the models can be re-identified using selected features by employing simple least squares to enhance the parameter estimates and consequently the model predictions.
 
In \cite{sun2020alven}, the authors have also used ideas of lifting and sparse optimization which has been applied on the actual nonlinear process data. However this approach requires to deal with the effect of unmeasured disturbances and measurement noise in the outputs. On the contrary our approach differs in the following aspects: (i) Development of local linear models across different operating points (sufficiently far) using traditional input excitation signals. In the present work, these models are assumed to be available. (ii) Simulation of noise-free data from multiple linear models (iii) Model fusion of multiple linear models utilizing the transformed simulation data and postulated nonlinear model, in combination with sparse optimization techniques (iv) Model re-identification post feature selection to enhance robustness. The major advantages of the proposed method in comparison to existing works is as follows:
\begin{itemize}
    \item Simplicity of experiment design (Step/PRBS) for development of multiple linear models.
    \item Reliability on multi-model simulation data in comparison to actual data enables effective rejection of measurement noise and unmeasured disturbance while developing the nonlinear model.
    \item Ease of nonlinear feature selection due to minimal noise in the features unlike the actual data.
\end{itemize} 

In the context of control, the closed-loop performance for a model-based control strategy such as model predictive control (MPC) heavily depends on the goodness of model predictions. Our proposed approach essentially develops a reliable nonlinear model applicable to a wider operating region, which can be easily deployed in a general nonlinear MPC framework. In a special case where none of the selected features have any kind of non-linearity wrt. inputs, a linear MPC can also be deployed.
 
To the best of the authors' knowledge, model fusion to obtain a global (applicable for relatively wider operating regions) nonlinear model using multiple local linear models has not been studied and considered in the literature. The applicability of the developed model is over a wider operating region, including the intermediate transit regions for which a local linear model is not available.  As a result, significant experimental efforts in the intermediate regions can be avoided for the sake of model development.

\section{Preliminaries and Problem formulation}
In this section, we review a variant of nonlinear auto-regressive with exogenous input (NARX) model structure, namely, the polynomial NARX (p-NARX) model. This model structure can accommodate mild nonlinear behaviour, which is adequate in many of the process applications. 
\subsection{Polynomial NARX model (\MakeLowercase{p}-NARX)}
The standard NARX model has the following form:
\begin{equation}
    y_k = f(y_{k-1}, y_{k-2}, \hdots, y_{k-l_y}, u_{k-l_d}, \hdots, u_{k-l_u-l_d})
\end{equation}
where $f(.)$ is some nonlinear function, while $l_y$, $l_u$ and $l_d$ represent the number of output lags (model order), input lags and the process delay, respectively. Restricting $f(.)$ to multi-variable polynomial model results in a linear-in-parameters form: $f(.) = \Psi(.)\alpha$, where $\Psi$ contains non-linear transformations of lagged inputs and outputs and $\alpha$ comprises of weights for all these features. This class of models are referred to as p-NARX. For instance, if $\Psi(.)$ contains $\{1, x, x^2\}$ as the polynomial basis functions, assuming $l_u =l_y = 2$ and $l_d = 0$, the resulting p-NARX model is given by,
\begin{equation}
\begin{aligned}
     y_{k} &= a_1y_{k-1}+a_2y_{k-2}+b_1u_{k-1}+b_2u_{k-2}+a_{11}y_{k-1}^2\\&+a_{22}y_{k-2}^2+b_{11}u_{k-1}^2+b_{22}u_{k-2}^2+a_{12}y_{k-1}y_{k-2}\\&+b_{12}u_{k-1}u_{k-2}+{ab}_{11}y_{k-1}u_{k-1}+{ab}_{22}y_{k-2}u_{k-2}\\&+{ab}_{12}y_{k-1}u_{k-2}+{ab}_{21}y_{k-2}u_{k-1}
\end{aligned}
\end{equation}
where $\alpha$ contains $a_{(.)}$ and $b_{(.)}$ coefficients.
\begin{note}\label{note:basis_function}
    The non-linear transformation basis function, namely, $\Psi(.)$ is not restricted to a polynomial basis. It can also accommodate other non-linear basis functions such as transcendental (logarithmic, exponential) and trigonometric functions. These can also be populated based on the underlying process knowledge. 
\end{note}

In this article, we restrict our focus to learning nonlinear polynomial NARX model (p-NARX) as these are known to adequately accommodate process non-linear behavior in wide applications. The basis function becomes $\{1, x, x^2, \hdots, x^q\}$, where $q \in \mathbb{Z}$. For instance, in \cite{sriniwas1995nonlinear} the authors have modelled the distillation column as p-NARX model considering upto $3^{rd}$-order interaction. Intuitively, for any given nonlinear function $f(.)$, first-order approximation using Taylor's series around an operating point can explain only linear dynamics whereas as we increase the order of approximation, the approximated function would be able to explain non-linear behavior as well. It is to be noted that all higher-order terms may not be necessary as we need only sufficient terms that can explain the underlying nonlinear behaviour over the operating region of interest. Despite truncation to a finite order, the number of terms for any given order would be extremely high due to the different polynomial combinations possible between the lagged inputs and outputs. 

Since we do not have any information on actual nonlinear features, we need to consider all potential features, which often results in over-fitting and multicollinearity issues. To address these, we add additional regularization terms on the parameters to the prediction error minization cost function while learning the model. It is well known that $l_{1}$ regularizer (LASSO) promotes sparsity-based solution, hence acts as feature selector (\cite{tibshirani1996regression}) while Ridge-regression ($l_2$ regularization) can take care of multicollinearity in the data. Therefore, we use Elastic-net (\cite{zou2005regularization}) as this can address the above-mentioned issues. Throughout this article the following assumptions are made as described below.

\begin{assumption}\label{assumption:1}
We assume the following:
\begin{enumerate}
    \item  Mild non-linear behaviour over the operating region of interest. 
    \item Local linear models are reliable locally around operating point.
\end{enumerate} 
\end{assumption}

The mild non-linear assumption is applicable to many real industrial processes which may comprise of sub-units such as tanks, reactors, columns etc. (\cite{xavier2024nonlinear}). In all process industries, the experiments are conducted locally around some operating point using step or PRBS test. This leads to reliable linear models locally around that operating point (see Assumption \ref{assumption:1}) under assumption of sufficient number of input-output samples. 

\subsection{Problem formulation}
Let us assume we have $l$-local linear models at distinct operating points satisfying Assumption \ref{assumption:1}. Further let the local linear model corresponding to $i^{th}$ operating point be denoted as $M_{i} \hspace{0.2cm} \forall i \in \{1, 2, \hdots, l\}$. Let $Y$ be vector containing outputs at various operating points, similarly let $Z$ be the input regressor matrix (containing linear and nonlinear features) and $\beta$ be the vector containing unknown co-efficients.
\begin{equation}
    Y := \begin{bmatrix} Y^{(1)} \\ Y^{(2)}\\ \vdots \\Y^{(l)} \end{bmatrix} \hspace{0.5cm} Z:= \begin{bmatrix} Z^{(1)} \\ Z^{(2)}\\ \vdots \\Z^{(l)} \end{bmatrix}
\end{equation}
Then the problem can be formulated as linear regression, 
\begin{equation}\label{eq:EN_optimization}
    \min_{\beta} \big\lVert Y - Z\beta \big\rVert_{2}^2 + \lambda \bigg( \gamma \lVert \beta\rVert_{1} + \frac{1-\gamma}{2} \lVert \beta\rVert_{2}^2 \bigg)
\end{equation}
where $\gamma \in (0,1)$ specifying the trade-off between LASSO and Ridge regression.
\begin{note}
    The local linear models are either already available (historical model) if not then step or PRBS tests can be performed on the real process plant to obtain these models.
\end{note}
\begin{problem}
    Efficient learning of nonlinear dynamic process (solving \eqref{eq:EN_optimization})  given linear models at different operating points.
\end{problem}
The efficiency in above statement means that we do not need to perform any additional experiment on the actual process plant (ie. without disturbing the process) to collect the input-output data.

\section{Model Fusion: Learning nonlinear dynamics}
In this section we propose a novel model fusion algorithm to estimates p-NARX model. The p-NARX model is an polynomial model wherein we have linear as well well as non-linear features. The major steps of the proposed algorithm is discussed in the next subsections.
\subsection{Linear features}
The local linear models are simulated to obtain noise-free input-output data corresponding to each operating point. Let the simulated input-output data for $i^{th}$ operating point is denoted by $\{u_j^{(i)}, y_{j}^{(i)}\}_{j=1}^N$ where $N$ is total number of samples, $u_j \in \mathbb{R}^m$ and $y_j \in \mathbb{R}^p$.
 Since the input-output lags is unknown therefore we assume $n_y > l_y$ and $n_u > l_u$. The total number of linear features is denoted by $n_l := n_y+n_u$. For simplicity of understanding we assume $m = p = 1$ ie. SISO case which can be easily extended for MIMO case.
 
\subsection{Appending non-linear features and feature selection}
In this article we have considered polynomial non-linearity (p-NARX model) so we make use of linear features obtained from simulated linear models to generate the polynomial non-linear features. Using basis function $\Psi$, the linear features can be transformed into non-linear features. Suppose we are interested upto quadratic non-linearity, then the number of possible non-linear features will be $n_{nl} = \frac{n_l(n_l+1)}{2}$. The regression data matrix (Z) has $n_{tot} = n_l + n_{nl} = \frac{n_{l}^2+3n_{l}}{2}$ features. Therefore the regression equation for $i^{th}$ operating point is given by:
\begin{equation}\begin{aligned}
     Y^{(i)} &= Z^{(i)}\beta
     &= \begin{bmatrix}
         Z_{l}^{(i)} & Z_{nl}^{(i)}
     \end{bmatrix}\beta
\end{aligned}
\end{equation}
where $Y^{(i)}\in \mathbb{R}^{N}$ is output data matrix corresponding to $i^{th}$ operating point, $Z^{(i)} \in \mathbb{R}^{N \times n_{tot}}$ is regressor matrix, $Z^{(i)}_{l} \in \mathbb{R}^{N \times n_l}$ is linear regressor matrix, $Z^{(i)}_{nl}\in \mathbb{R}^{N \times n_l}$ is non-linear regressor matrix and $\beta \in \mathbb{R}^{n_{tot}}$ is co-efficient matrix. Stacking together for all operating points we get:
\begin{equation}
    \underbrace{\begin{bmatrix} Y^{(1)} \\ Y^{(2)}\\ \vdots \\Y^{(l)} \end{bmatrix}}_{=: Y} = \underbrace{\begin{bmatrix} Z^{(1)} \\ Z^{(2)}\\ \vdots \\Z^{(l)} \end{bmatrix}}_{=: Z} \beta
\end{equation}
In compact form we can write above equation as $Y =Z\beta$ where $Y \in \mathbb{R}^{lN}$ and $Z \in \mathbb{R}^{lN \times n_{tot}}$. Since the features affecting the dynamics of the system is unknown so we have taken all possible combination of the features which may lead to over-fitting and data colinearity issues. To discover the important and independent features we use Elastic net as regularization method. 
Then the problem can be formulated as linear regression, 
\begin{equation}\label{eq:EN_optimization_final}
    \min_{\beta} \big\lVert Y - Z\beta \big\rVert_{2}^2 + \lambda \bigg( \gamma \lVert \beta\rVert_{1} + \frac{1-\gamma}{2} \lVert \beta\rVert_{2}^2 \bigg)
\end{equation}
where $\gamma \in (0,1)$ specifying the trade-off between LASSO and Ridge regression. Let $n_s$ be the number of selected features by solving the above problem and let $\hat{\beta_s} \in \mathbb{R}^{n_s}$ denotes the co-efficient of selected features. A straightforward way would be to use solution of above equation $\hat{\beta}_s$ as final estimate. But it turns-out that due to the regularization term, the actual importance of the co-efficient gets re-distributed amongst other co-efficients (features) \cite{}. Therefore it is good practice to perform model re-identification post feature selection\cite{}. 
\subsection{Model re-identification and validation}
The next step is to perform model re-identification using selected features. Let $Z_{f} \in \mathbb{R}^{lN \times n_s}$ denotes the final regessor matrix corresponding to the selected features and $\beta_{f} \in \mathbb{R}^{n_s}$ denotes the final re-identified co-efficient vector corresponding to the selected features. The model re-identification can be done by solving following equation:
\begin{equation}
    \min_{\beta_f} \big\lVert Y - Z_f\beta_f \big\rVert_{2}^2
\end{equation}
The above problem is solved using p-NARX method to obtain the final model. Next we perform model validation on test dataset. We use prediction error as the validation metric given by:
\begin{equation}
    MSE = \frac{1}{N_{v}} \sum_{k=1}^{N_{v}}\big(y_{k} - \hat{y}_{k}\big)^2
\end{equation}
where $N_v$ is number of samples in test data, $y_{(.)}$ is the actual output and $\hat{y}_{(.)}$ is the predicted output.  The steps involved in the proposed algorithm is presented in Algorithm \ref{algo:model_fusion}.

\begin{algorithm}[h]
    \caption{Model fusion (p-NARX model)}\label{algo:model_fusion}
    \textbf{Input:} Input-output data at various operating points ie. $\{u_j^{(i)}, y_{j}^{(i)}\}_{j=1}^N$ $\forall i \in \{1, 2, \hdots, l\}$.\\
    \textbf{Output:} Global nonlinear p-NARX model.
    \begin{enumerate}
        \item Choose $n_y > l_y$ and $n_u > l_u$.
        \item Generate linear features using given data with $n_y$ and $n_u$ lags and formulate $Z_l$ matrix.
        \item Using linear features generate all possible polynomial (nonlinear) features and formulate $Z_{nl}$ matrix.
        \item Append linear and nonlinear features together and choose hyperparameters $\lambda$ and $\gamma$ for feature selection.
        \item Solve equation \eqref{eq:EN_optimization_final} for feature selection.
        \item Perform model re-identification to obtain p-NARX model using selected features.
        \item Validate model using test data (unseen by the model during training).
    \end{enumerate}
\end{algorithm}


\section{Case Studies}
In this section we verify the effectiveness of the proposed algorithm by identifying several nonlinear dynamical systems. For simplicity we have considered only SISO systems which can be easily extended for MIMO case. In all cases, the comparison was done with model predictions from available linear model at the nearest operating points. 
\subsection{Toy example}
\begin{figure}[h]
\centering
\includegraphics[width=8cm]{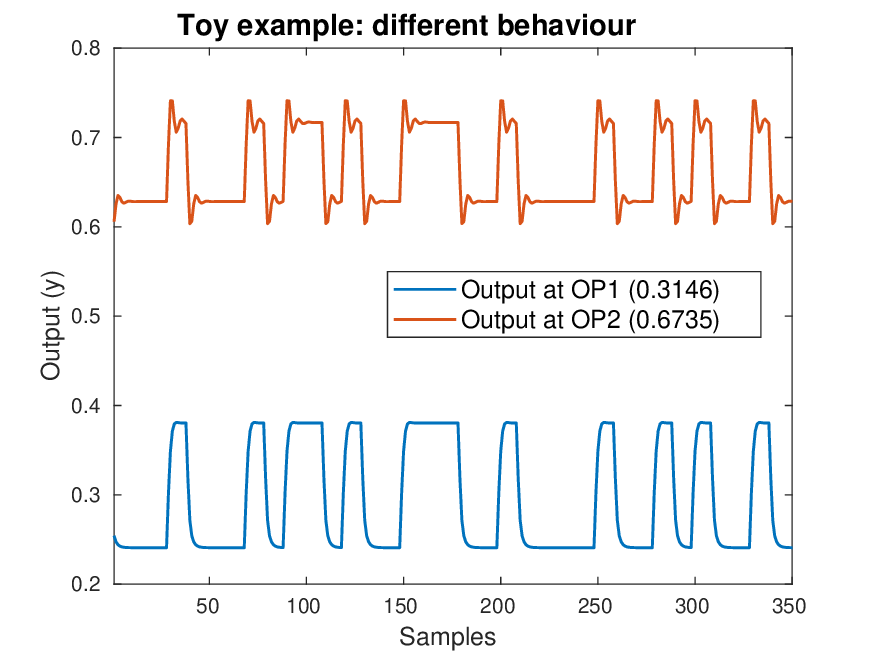}
\caption{Toy example: Different behaviour at two operating points}
\label{fig:toy_example_behaviour}
\end{figure}

Let us consider nonlinear polynomial model with quadratic non-linearity given by:
\begin{equation*}\begin{aligned}
        y(k) &= 0.5y(k-1)+0.25y(k-2) - 0.5y^2(k-2)\\ 
        &+ 0.1y(k-1)u(k-1)+u(k-1)+0.25u(k-2)
\end{aligned}
\end{equation*}
The local linear models were given at two ($l=2$) steady-sate operating points namely $OP1 - (0.1, 0.3146)$ and $OP2 - (0.3, 0.6735)$. Let the models at given operating points are denoted by $M_{1}$ and $M_2$ respectively. From Fig. \ref{fig:toy_example_behaviour}, it is clear that around operating point 1 $(OP1)$ ie. $0.3146$ model response is similar to first-order while at operating point 2 $(OP2)$ ie. $0.6735$ the model behaviour is similar to second-order dynamics. So the model is exhibiting different behaviour at two different operating points. Intuitively, using one model to predict behaviour at in-between operating point may lead to inferior prediction. Hence we use our proposed algorithm (Algorithm \ref{algo:model_fusion}) to estimate a reliable nonlinear model.

\begin{figure}[h]
\centering
\includegraphics[width=8cm]{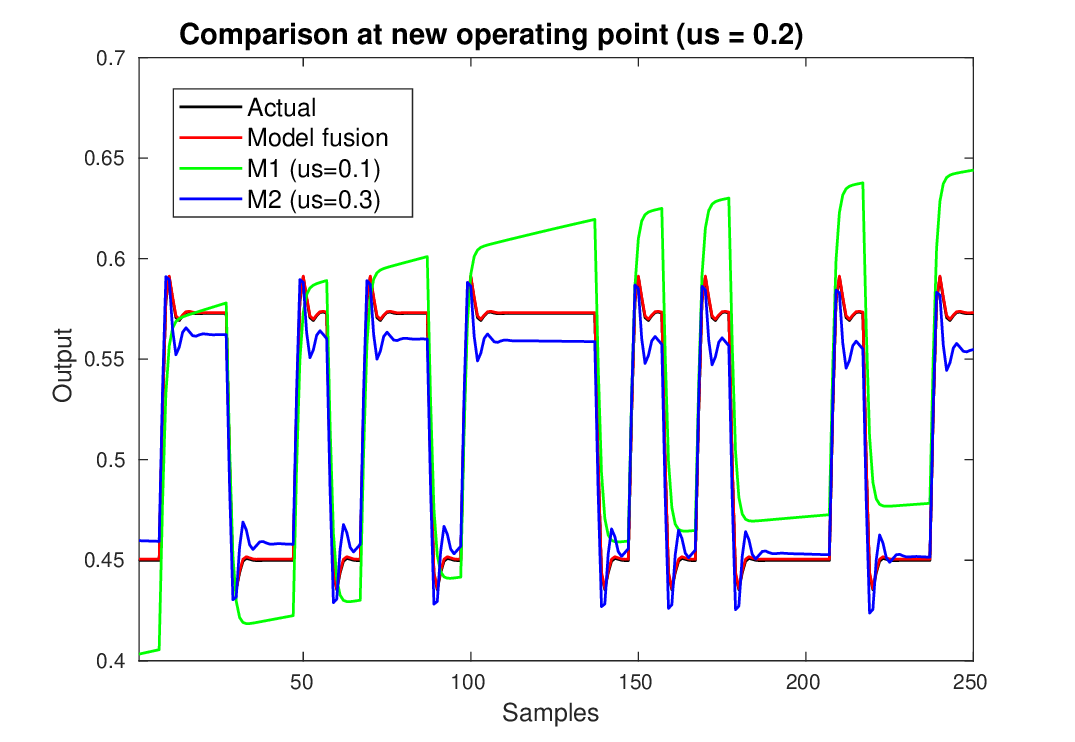}
\caption{Comparison of model predictions at new operating point (us = 0.2)}
\label{fig:toy_example_usinter}
\end{figure}

Since the model order is not known we chose $n_u = 3 > l_u$, $n_y =3 > l_y$. Therefore, total features $n_{tot} = 27$. Total number of training samples for each operating point is $N = 448$ and number of test data samples $N_v = 155$. The regressor matrix is of size $Z \in \mathbb{R}^{698 \times 27}$. The hyper-parameters chosen for this case are $\gamma  = 0.5$ (equal weightage to address over-fitting and colinearity) while $\lambda$ was chosen using 3-fold cross validation (see \eqref{eq:EN_optimization_final}). The optimization problem given by equation \eqref{eq:EN_optimization_final}, selected nine features ie. $n_s = 9$. The model was re-identified as global p-NARX model using the selected features and then validated using test data (see Algorithm \ref{algo:model_fusion}). In order to show the efficiency of the proposed method we chose an intermediate steady-state operating point ie. $(0.2, 0.5214)$. The model predictions were plotted for (a) given linear models at $OP1$ and $OP2$ (b) global p-NARX model - proposed method. From Fig. \ref{fig:toy_example_usinter} we can deduce that neither of the linear models ($M_1$ and $M_2$) have good prediction capability at intermediate operating point $(0.2, 0.5214)$ while the proposed model fusion approach works very well. The MSE at intermediate operating point for proposed algorithm was found to be more than 60 times lesser than that of linear model predictions (see Table \ref{tab:toy_example_MSE}). To further investigate the effectiveness of the proposed method we compare the model predictions at various operating points. From Table \ref{tab:toy_example_MSE}, it can be deduced that the proposed model fusion approach works better than individual linear model predictions.

\begin{table}[h]
\begin{center}
				\caption{Performance comparison for toy example at various operating points}
				\label{tab:toy_example_MSE}
\scalebox{0.8}{\begin{tabular}{|c|c|c|c|c|c|}
\hline
\textbf{\begin{tabular}[c]{@{}c@{}}OP \\ (us)\end{tabular}} & \textbf{\begin{tabular}[c]{@{}c@{}}MSE\\ (Model fusion)\end{tabular}} & \textbf{\begin{tabular}[c]{@{}c@{}}MSE\\ ($M_1$)\end{tabular}} & \textbf{\begin{tabular}[c]{@{}c@{}}MSE\\ ($M_2$)\end{tabular}} & \textbf{\begin{tabular}[c]{@{}c@{}}MSE ratio\\ ($M_1$/$MF$)\end{tabular}} & \textbf{\begin{tabular}[c]{@{}c@{}}MSE ratio\\ ($M_2$/$MF$)\end{tabular}} \\ \hline
0.05                                                                                  & 4.237e-06                                                                  & 5.41e-04                                                         & 0.0026                                                         & 127.66                                                               & 619.95                                                              \\ \hline
0.1                                                                                 & 2.537e-06                                                                & 3.286e-05                                                       & 0.0012                                                        & 12.95                                                    & 49.29                                                             \\ \hline
0.2                                                                                & 9.839e-06                                                               & 0.0015                                                         & 6.65e-04                                                         & 154.84                                                              & 67.6                                                             \\ \hline
0.3                                                                                & 2.969e-05                                                                & 0.0069                                                         & 5.779e-04                                                        & 232.96                                                              & 19.46                                                               \\ \hline
0.35                                                                                 & 4.855e-05                                                                & 0.0115                                                         & 5.77e-04                                                      & 237.1                                                              & 11.88                                                      \\ \hline
\end{tabular}}
\end{center}
\end{table}
\subsection{Conical tank process}
Conical tanks find wide applications in various process industries, including hydro-metallurgical, food processing, and wastewater treatment industries. Moreover understanding the nonlinear dynamics for conical tank can provide valuable insights for comprehending other types of nonlinear processes in process industries. 

Dynamics of nonlinear conical tank system is given by (\cite{vatsal2020multi}):
\begin{equation*}
    \frac{dh}{dt} = \frac{\alpha(q_i - C_d \sqrt{h})}{h^2}
\end{equation*}
where $\alpha = \frac{4H^2}{\pi D^2}$ is constant, $h$ denotes water level (height), $q_i$ is inlet flow-rate, $C_d$ is valve constant, $D$ is maximum diameter and $H$ is maximum height. 

\begin{figure}[h]
\centering
\includegraphics[width=8cm]{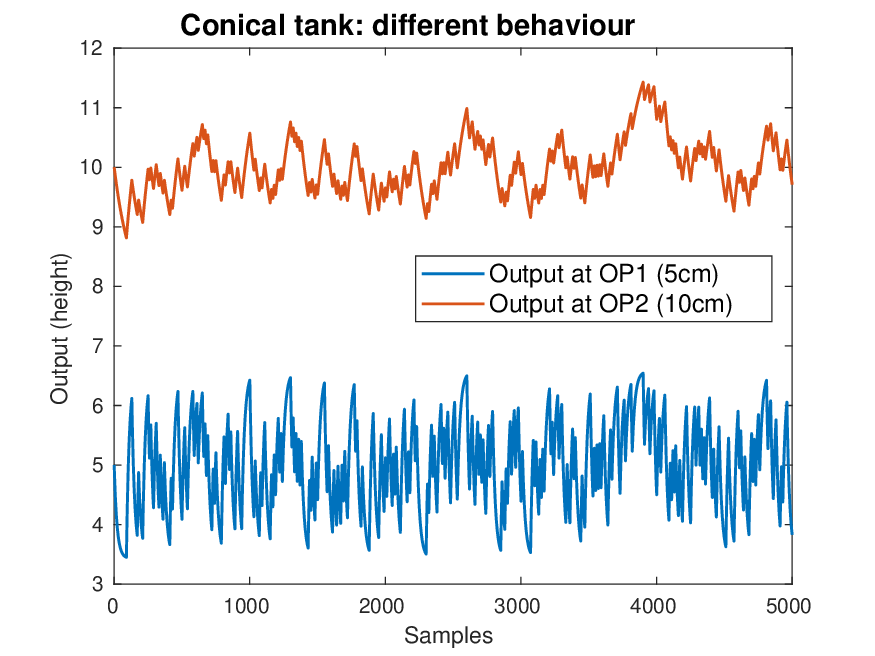}
\caption{Conical tank: Different behaviour at two operating points}
\label{fig:conical_tank_behaviour}
\end{figure}
\begin{figure}[h]
\centering
\includegraphics[width=8cm]{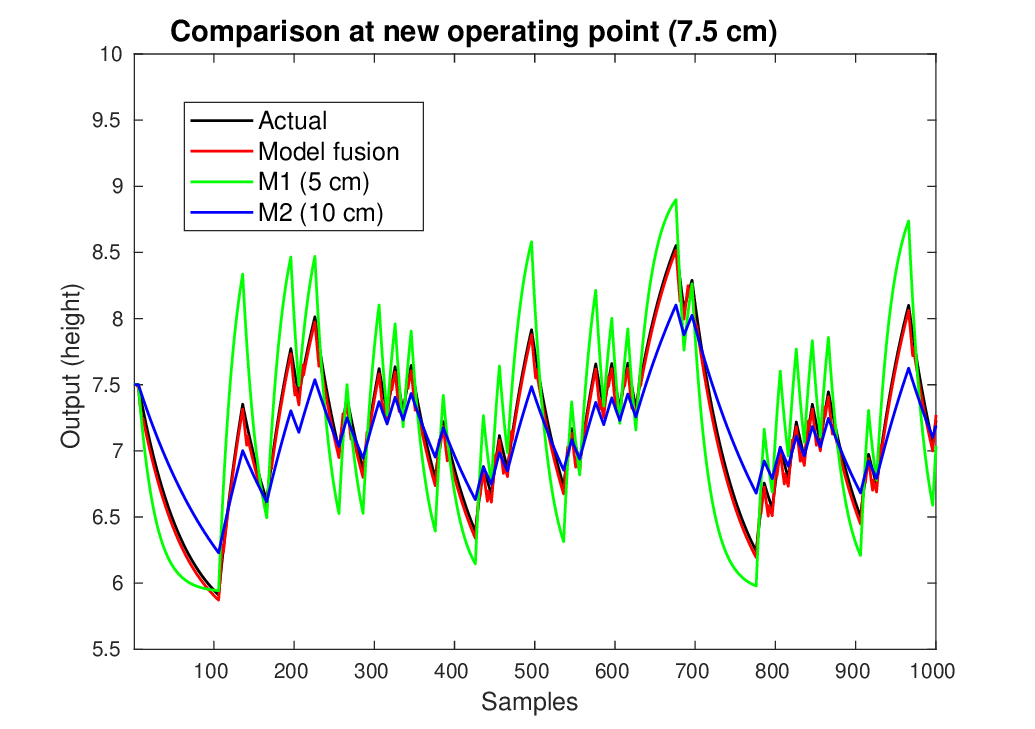}
\caption{Comparison of model predictions at new operating point (7.5 cm)}
\label{fig:conical_tank_7p5cm}
\end{figure}

In this case the local linear models were given at two ($l=2$) operating points which is steady-state height ie. $h_s = 5 cm$ and $h_s = 10 cm$. Similar to previous case of toy example, the model behaviour is different at different operating point (refer Fig. \ref{fig:conical_tank_behaviour}). At operating point 1 (OP1) ie. $h_s = 5 cm$ the dynamics is faster as compared to OP2 ie. $h_s = 10 cm$ (see Fig. \ref{fig:conical_tank_behaviour}). Similar process was followed as in toy example. The parameters used in this case are $D = 30 cm$, $H = 62 cm$, $C_d = 1$, $n_u =n_y =3$, $n_{tot} = 27$, $N = 7109$, $N_v = 3048$, $\alpha = 0.75$ and $\lambda$ was chosen based on 5-fold cross-validation. The optimization problem selected 12 features ie. $n_s = 12$. Model re-identification to obtain p-NARX model was done using selected features and validation was performed using test data. In order to show the efficiency of the proposed method we chose an intermediate operating point ie. $h_s = 7.5 cm$. Model predictions were plotted for (a) linear models at given operating point and (b) global p-NARX model fusion (proposed method). From Fig. \ref{fig:conical_tank_7p5cm} we can deduce that neither of the linear models ($M_1$ and $M_2$) have good prediction capability at intermediate operating point $h_s = 7.5cm$ while the proposed model fusion approach works very well. From Table \ref{fig:conical_tank_behaviour}, MSE at the intermediate operating point for the proposed algorithm is atleast 10 times better as compared to the individual linear models. To further investigate the effectiveness of the proposed method we compare the model predictions at various operating points (see Table \ref{tab:canonical_tank_MSE}), it can be deduced that model fusion approach works better than individual linear model predictions. Moreover, the performance at operating points where local models were given was found to be better than that of the proposed method. This is intuitive because linear models are very good approximation at operating points where it is developed, hence better performance.

\begin{table}[h]
\begin{center}
				\caption{Performance comparison for conical tank at various operating points}
				\label{tab:canonical_tank_MSE}
\scalebox{0.8}{\begin{tabular}{|c|c|c|c|c|c|}
\hline
\textbf{\begin{tabular}[c]{@{}c@{}}OP \\ ($h_s$)\end{tabular}} & \textbf{\begin{tabular}[c]{@{}c@{}}MSE\\ (Model fusion)\end{tabular}} & \textbf{\begin{tabular}[c]{@{}c@{}}MSE\\ ($M_1$)\end{tabular}} & \textbf{\begin{tabular}[c]{@{}c@{}}MSE\\ ($M_2$)\end{tabular}} & \textbf{\begin{tabular}[c]{@{}c@{}}MSE ratio\\ ($M_1$/$MF$)\end{tabular}} & \textbf{\begin{tabular}[c]{@{}c@{}}MSE ratio\\ ($M_2$/$MF$)\end{tabular}} \\ \hline
4                                                                                  & 0.0336                                                                & 0.0495                                                         & 0.4279                                                         & 1.4732                                                               & 12.7351                                                              \\ \hline
5                                                                                  & 0.0061                                                                & 5.36e-06                                                       & 0.2561                                                         & \textbf{8.79e-04}                                                    & 41.9836                                                              \\ \hline
7.5                                                                                & 0.0042                                                                & 0.1140                                                         & 0.0443                                                         & 27.1429                                                              & 10.5476                                                              \\ \hline
8.5                                                                                & 0.0058                                                                & 0.1728                                                         & 0.0144                                                         & 29.7931                                                              & 2.4828                                                               \\ \hline
10                                                                                 & 0.0082                                                                & 0.2477                                                         & 1.143e-04                                                      & 30.2073                                                              & \textbf{0.0139}                                                      \\ \hline
11                                                                                 & 0.0147                                                                & 0.2881                                                         & 0.0027                                                         & 19.5989                                                              & 0.1837                                                               \\ \hline
\end{tabular}}
\end{center}
\end{table}
\subsection{Hammerstein-Wiener models}
\begin{figure}[h]
\centering
\includegraphics[width=8cm]{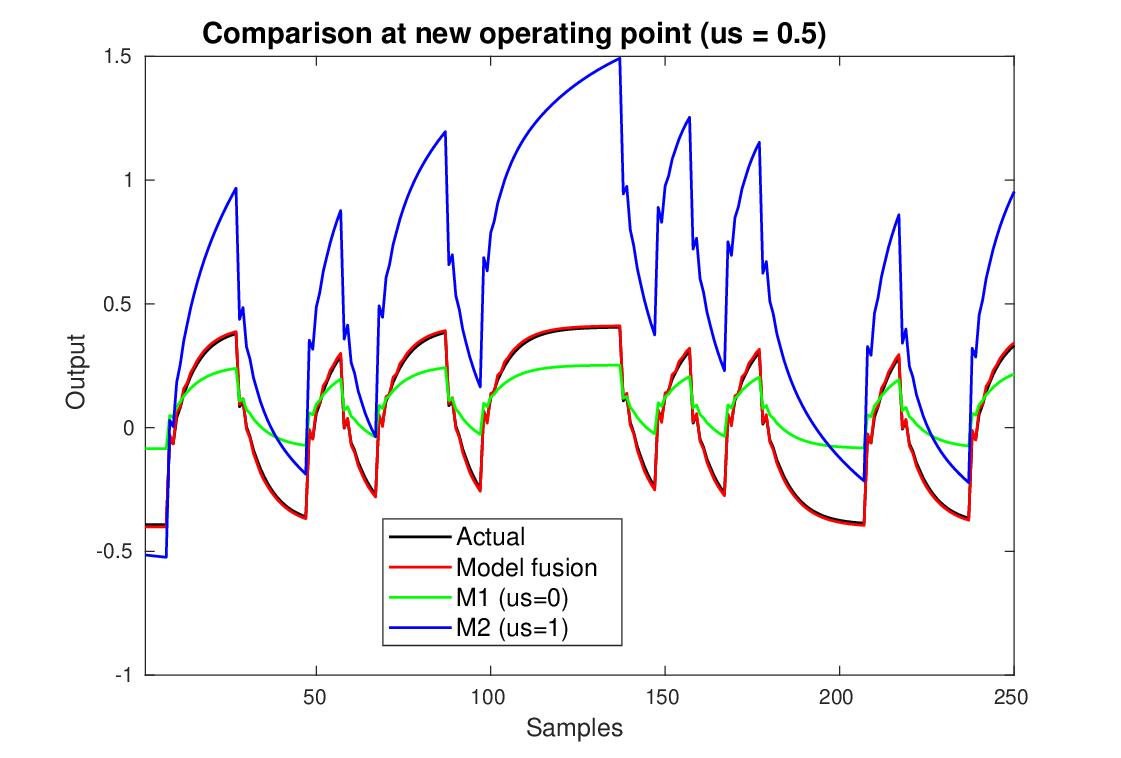}
\caption{Comparison of model predictions at new operating point (us = 0.5)}
\label{fig:hammerstein_example}
\end{figure}

Hammerstein-Wiener models is represented by a linear transfer function and capture the nonlinearities using nonlinear functions of inputs and outputs of the linear system. The configuration looks like series connection of static nonlinear blocks with a dynamic linear block. It has several applications such as modeling electromechanical system (\cite{Schoukens2012HammwesteinWiener_identification}) and radio frequency components (\cite{taringou2010behaviour}), predictive control of chemical processes like Bio-reactors (\cite{luo2018HW_bioreactor}), etc. Let us consider the following example:

\begin{equation*}\begin{aligned}
& s(t) = \frac{A(z)}{B(z)}x(t) + v(t)\\
    &A(z) = 1 - 0.45z^{-1} -0.35z^{-2}\\
    &B(z) = 0.5z^{-1} - 0.25z^{-2}\\
    &x(t) = f(u(t)) =  0.5 u(t) - 0.18 u^2(t)\\
    &y(t) = g(s(t)) = \frac{1}{0.3} \bigg(-1+ \sqrt{1+0.6s(t)}\bigg) 
\end{aligned}
\end{equation*}
where $f(.)$ and $g(.)$ are static nonlinear function acting on input ($u(t)$) and intermediate output ($s(t)$) respectively. In this case, local linear models were given at two ($l=2$) operating points ie. $M_1 (u_s = 0)$ and $M_2 (u_s = 1)$. Similar analysis was performed as in previous examples with $n_u = n_y = 3$. From Fig. \ref{fig:hammerstein_example} it can be clearly deduced that individual linear model predictions are worse than the proposed model fusion approach.
\section{Conclusion}
We proposed a practical and efficient model fusion approach from multiple linear models to obtain a global non-linear reliable model over a wider operating region. The efficacy of the proposed approach has been demonstrated in three simulation examples. The predictions of the developed nonlinear models are observed to be close to the real process responses even for the intermediate operating regions for which there is no local linear model available without the requirement of additional experimental efforts. The results show promising steps towards developing models in unknown transition regions which is necessary in applications involving frequent random grade changes over wider operating regions and also in ensuring bump-less transfer while switching from one grade to the another. Future work considers an extension of these ideas for multi-input multi-output systems, accounting for variable delays and model fusion in state space framework.




\bibliography{Final_version}             

\begin{thebibliography}{18}
\providecommand{\natexlab}[1]{#1}
\providecommand{\url}[1]{\texttt{#1}}
\providecommand{\urlprefix}{URL }
\expandafter\ifx\csname urlstyle\endcsname\relax
  \providecommand{\doi}[1]{doi:\discretionary{}{}{}#1}\else
  \providecommand{\doi}{doi:\discretionary{}{}{}\begingroup \urlstyle{rm}\Url}\fi

\bibitem[{Adeniran and El~Ferik(2016)}]{adeniran2016modeling}
Adeniran, A.A. and El~Ferik, S. (2016).
\newblock Modeling and identification of nonlinear systems: A review of the multimodel approach—part 1.
\newblock \emph{IEEE Transactions on Systems, Man, and Cybernetics: Systems}, 47(7), 1149--1159.

\bibitem[{Johansen and Foss(1999)}]{johansen1999multiple}
Johansen, T.A. and Foss, B.A. (1999).
\newblock Multiple model approaches to modelling and control.
\newblock \emph{International journal of control}, 72(7-8), 575--575.

\bibitem[{Ljung(1998)}]{ljung1998system}
Ljung, L. (1998).
\newblock System identification.
\newblock In \emph{Signal analysis and prediction}, 163--173. Springer.

\bibitem[{Ljung(2010)}]{ljung2010perspectives}
Ljung, L. (2010).
\newblock Perspectives on system identification.
\newblock \emph{Annual Reviews in Control}, 34(1), 1--12.

\bibitem[{Luo and Song(2018)}]{luo2018HW_bioreactor}
Luo, X.S. and Song, Y.D. (2018).
\newblock Data-driven predictive control of hammerstein--wiener systems based on subspace identification.
\newblock \emph{Information Sciences}, 422, 447--461.

\bibitem[{Nikolakopoulou et~al.(2020)Nikolakopoulou, Hong, and Braatz}]{nikolakopoulou2020feedback}
Nikolakopoulou, A., Hong, M.S., and Braatz, R.D. (2020).
\newblock Feedback control of dynamic artificial neural networks using linear matrix inequalities.
\newblock In \emph{2020 59th IEEE Conference on Decision and Control (CDC)}, 2210--2215. IEEE.

\bibitem[{Paduart et~al.(2012)Paduart, Lauwers, Pintelon, and Schoukens}]{Schoukens2012HammwesteinWiener_identification}
Paduart, J., Lauwers, L., Pintelon, R., and Schoukens, J. (2012).
\newblock Identification of a wiener--hammerstein system using the polynomial nonlinear state space approach.
\newblock \emph{Control Engineering Practice}, 20(11), 1133--1139.

\bibitem[{Pillonetto et~al.(2023)Pillonetto, Aravkin, Gedon, Ljung, Ribeiro, and Sch{\"o}n}]{pillonetto2023deepnetwrok}
Pillonetto, G., Aravkin, A., Gedon, D., Ljung, L., Ribeiro, A.H., and Sch{\"o}n, T.B. (2023).
\newblock Deep networks for system identification: a survey.
\newblock \emph{arXiv preprint arXiv:2301.12832}.

\bibitem[{Prasad et~al.(2020)Prasad, Kedia, and Rao}]{vatsal2020multi}
Prasad, G.M., Kedia, V., and Rao, A.S. (2020).
\newblock Multi-model predictive control (mmpc) for non-linear systems with time delay: an experimental investigation.
\newblock In \emph{2020 First IEEE International Conference on Measurement, Instrumentation, Control and Automation (ICMICA)}, 1--5. IEEE.

\bibitem[{Ren et~al.(2022)Ren, Alhajeri, Luo, Chen, Abdullah, Wu, and Christofides}]{ren2022mltutorial}
Ren, Y.M., Alhajeri, M.S., Luo, J., Chen, S., Abdullah, F., Wu, Z., and Christofides, P.D. (2022).
\newblock A tutorial review of neural network modeling approaches for model predictive control.
\newblock \emph{Computers \& Chemical Engineering}, 107956.

\bibitem[{Schoukens and Ljung(2019)}]{schoukens2019nonlinear}
Schoukens, J. and Ljung, L. (2019).
\newblock Nonlinear system identification: A user-oriented road map.
\newblock \emph{IEEE Control Systems Magazine}, 39(6), 28--99.

\bibitem[{Sriniwas et~al.(1995)Sriniwas, Arkun, Chien, and Ogunnaike}]{sriniwas1995nonlinear}
Sriniwas, G.R., Arkun, Y., Chien, I.L., and Ogunnaike, B.A. (1995).
\newblock Nonlinear identification and control of a high-purity distillation column: a case study.
\newblock \emph{Journal of Process Control}, 5(3), 149--162.

\bibitem[{Sun and Braatz(2020)}]{sun2020alven}
Sun, W. and Braatz, R.D. (2020).
\newblock Alven: Algebraic learning via elastic net for static and dynamic nonlinear model identification.
\newblock \emph{Computers \& Chemical Engineering}, 143, 107103.

\bibitem[{Taringou et~al.(2010)Taringou, Hammi, Srinivasan, Malhame, and Ghannouchi}]{taringou2010behaviour}
Taringou, F., Hammi, O., Srinivasan, B., Malhame, R., and Ghannouchi, F.M. (2010).
\newblock Behaviour modelling of wideband rf transmitters using hammerstein--wiener models.
\newblock \emph{IET circuits, devices \& systems}, 4(4), 282--290.

\bibitem[{Tibshirani(1996)}]{tibshirani1996regression}
Tibshirani, R. (1996).
\newblock Regression shrinkage and selection via the lasso.
\newblock \emph{Journal of the Royal Statistical Society Series B: Statistical Methodology}, 58(1), 267--288.

\bibitem[{Xavier et~al.(2024)Xavier, Patnaik, and Panda}]{xavier2024nonlinear}
Xavier, J., Patnaik, S., and Panda, R.C. (2024).
\newblock Nonlinear system identification in coherence with nonlinearity measure for dynamic physical systems—case studies.
\newblock \emph{Nonlinear Dynamics}, 1--27.

\bibitem[{Zhu(2001)}]{zhu2001multivariable}
Zhu, Y. (2001).
\newblock \emph{Multivariable system identification for process control}.
\newblock Elsevier.

\bibitem[{Zou and Hastie(2005)}]{zou2005regularization}
Zou, H. and Hastie, T. (2005).
\newblock Regularization and variable selection via the elastic net.
\newblock \emph{Journal of the Royal Statistical Society Series B: Statistical Methodology}, 67(2), 301--320.

\end{thebibliography}
                                                   







\end{document}